\documentclass{PoS}

\usepackage{epsf}
\usepackage{amssymb}
\usepackage{amsmath}
\usepackage{amsfonts}
\usepackage{cite}
\usepackage[small]{caption2}

\usepackage[T1]{fontenc}
\usepackage{psfrag,epsfig,graphicx,graphics}

\newcommand{\be}{\begin{equation}}
\newcommand{\beq}{\begin{equation}}
\newcommand{\ee}{\end{equation}}
\newcommand{\eq}{\end{equation}}
\newcommand{\eeq}{\end{equation}}

\newcommand{\bea}{\begin{eqnarray}}
\newcommand{\eea}{\end{eqnarray}}

\def\slashchar#1{\setbox0=\hbox{$#1$}
   \dimen0=\wd0
   \setbox1=\hbox{/} \dimen1=\wd1
   \ifdim\dimen0>\dimen1
      \rlap{\hbox to \dimen0{\hfil/\hfil}}
      #1
   \else
      \rlap{\hbox to \dimen1{\hfil$#1$\hfil}}
      /gdatdafinal2.tex
   \fi}

\def\h#1{\widehat{#1}}

\def\px#1#2{\frac{\partial#1}{\partial#2}}

\def\C{\mathcal{C}}

\def\F{\mathcal{F}}

\def\V{\mathcal{V}}

\def\NB{\mathbb{N}}

\def\stmath#1{#1}
\def\structure#1{#1}
\def\aut#1{#1}
\def\alert#1{#1}
\def\violet#1{#1}

\title{Virtual Compton Scattering off a Spinless Target in the AdS/QCD
correspondence}

\ShortTitle{Virtual Compton Scattering off a Spinless Target in the AdS/QCD
correspondence}

\author{C.~Marquet%
\\
Physics Department, Theory Unit, CERN, 1211 Gen\`eve 23, Switzerland\\
        E-mail: \email{cyrille.marquet@cern.ch}}

\author{C.~Roiesnel
\\
 Centre de Physique Th\'eorique, \'Ecole Polytechnique,
   CNRS,
   91128 Palaiseau, France\\
E-mail: \email{Claude.Roiesnel@cpht.polytechnique.fr}}

\author{\speaker{S.~Wallon}
\\
LPT, Universit{\'e} Paris-Sud, CNRS, 91405 Orsay, France \ {\em \&} \\
UPMC Univ. Paris 06, facult\'e de physique, 4 place Jussieu, 75252 Paris Cedex  05, France\\
        E-mail: \email{wallon@th.u-psud.fr}}

\abstract{ We perform a study of the doubly virtual Compton scattering off a spinless target
  $\gamma^*P\to\gamma^*P'$ within the Anti-de Sitter(AdS)/QCD
  formalism.  We find that the general structure allowed by the
  Lorentz invariance and gauge invariance of the Compton amplitude is
  not easily reproduced with the standard recipes of the AdS/QCD
  correspondence.  In the soft-photon regime, where the semi-classical
  approximation is supposed to apply best, we show that the
  measurements of the electric and magnetic polarizabilities of a
  target like the charged pion in real Compton scattering, can already
  serve as stringent tests, and presumably exclude results based on the AdS/QCD
correspondence in its minimal version.
}

\FullConference{35th International Conference of High Energy Physics\\
                 July 22-28, 2010\\
                 Paris, France}

\begin{document}

\section{Introduction}
\label{Sec_Int}


The AdS/CFT correspondence \cite{MA98}
postulates  a relation \cite{WI98, GKP98} between weakly coupled string
theories living in the bulk of an anti-de Sitter (AdS) space and
strongly coupled conformally invariant field theories defined on its
boundary, based on the formal identification of the conformal group
 of the 4-d Minkowski space and of the
isometry group of AdS$_5$. 
The efforts within the so-called AdS/QCD approach rely on
the assumption that the AdS/CFT dictionary can still describe the
strong coupling regime of a confining gauge theory like QCD, despite
the breaking of conformal invariance by radiative corrections.

We here report on recent results obtained in the study of the Virtual Compton scattering off a spinless target \cite{us}. We show that the obtained amplitude does not satisfy the expected scaling for large photon virtualities. At low energy, the obtained electric and magnetic polarizabilities vanish, in contradiction with the latest experimental results.

\section{Gauge/Gravity duality}
\label{sec:Gauge/Gravity}

The AdS/CFT correspondence conjectures a relation between a IIB string theory compactified on $AdS_5 \times S^5$ and a 
${\cal N}=4$ super-Yang-Mills theory. 
The operator in CFT
${\cal O}(\vec x)$  creating asymptotic states is coupled to
the source
$\phi_0(\vec x)$, which is the boundary value of the  bulk field $\phi(\vec x, z)$ \footnote{We denote by $\vec{x}$ (or just $x$) the 4d-vector, and $z$ the holographic direction.}.
Both generating functions in CFT and AdS are related through: 
\beq
\label{dualityZ}
Z_{CFT}[\phi_0]=\langle e^{\int d^4 x \, \phi_0(\vec x) \, {\cal O}(\vec x) } \rangle_{CFT}
= Z_{string} 
\Bigg[ \phi(\vec x, z)\Big|_{z = z_{min}} = \phi_0(\vec x)
\Bigg]\,.
\eq
The 5d metric with Lorentz group $SO(1,3)$  as isometry subgroup reads (with $\eta\! =\! \text{diag}(-1,1,1,1,1))$ 
\beq
   ds^2 = a^2(z)\left(dx_{\mu}dx^{\mu}+dz^2\right)\,,\quad
  g^{mn} = g_{mn}^{-1} = a^{-2}(z)\eta^{mn}\,,
  \quad m=0,1,2,3,4\,.
\eq
In the $AdS_5$ metric, $a(z)=R/z$ (negative curvature, corresponding to a negative cosmological constant),  $\sqrt{-g}=R^5/z^5$ \,($R=1$ from now on).
The UV limit $z \to z_{min}$ is the CFT boundary.
The field content is made of
a 5d vector $U(1)$ field $A_m(x,z),$ dual to the electromagnetic current,
and 
a 5d massive scalar field $\Phi(x,z)$ dual to an operator creating the spinless target.

In the large $N$ and large 't Hooft coupling $\lambda= g_{YM}^2 N$
limit, the IIB string theory reduces to a IIB SUGRA theory in 10 dimensions, when
$R^4/l_s^4 \sim g_s \, N \sim g_{YM}^2 N \gg 1$, 
$R$ being the $AdS_5$ and $S^5$ radius and
$l_s$ the string length. In that limit
$g_s \ll 1$ and $R \gg l_s$ the string theory reduces to classical SUGRA with decoupling of massive string excitations.

The conformal invariance breaking due to QCD confinement  can be introduced by an IR cut-off at $z=1/\Lambda$
\cite{PS01} (``hard-wall`` model) or by a background dilaton field
$\chi$ \cite{KKSS} (``soft-wall`` model),   used here\footnote{Both approaches reproduce qualitatively the spectra of low lying hadron states and various decay and coupling constants \cite{spectrum}.}.
The action for  $\Phi$ in this $\chi$ background \nolinebreak reads
\beq
\label{free_action}
  S_{\Phi} = \frac{1}{2}\int
  d^4x\,dz \, \sqrt{-g}\, e^{-\chi}\left(g^{ij}\partial_i\Phi\partial_j\Phi +
    m^2_S\Phi^2\right)\,.
\eq
\nopagebreak
The
classical free field equation  
exhibits 
 plane-wave solutions in 4d space:
$
  \Phi^{(0)}(x,z)
=
  e^{ip\cdot x}\,\hat{\Phi}(z)
$  \pagebreak
where $\{\hat{\Phi}_n(z)\,,\
n\in\NB\}$ is a complete set 
 of  Kaluza-Klein (KK) solutions.
The Fourier transform  of  the obtained
 Green function is a bulk to bulk propagator 
with an explicit  K\"allen-Lehman spectral  representation  when summing over  the  KK excitations: 
$$
  \h{G}(z,z';p) = -\sum_{n=0}^{\infty}
  \frac{\h{\Phi}^{\star}_n(z)\h{\Phi}_n(z')}{p^2+m^2_n-i\epsilon} \,.
$$
The action of a classical solution leaves only the boundary term. 
The classical partition function
 \beq
 Z_{5D}[\Phi_S] = \int_{\Phi(x,z_{\text{min}})=\Phi_S(x)}\,e^{iS[\Phi]}
  \quad\propto\quad e^{iS_{\text{cl}}[\Phi_S]} \,,
 \eq
 combined with the holographic identification of two-point functions from AdS to CFT leads to
\beq
\hspace{-.05cm} \langle O(x)O(x') \rangle \!\!=\! \!\frac{1}{Z[\Phi_S]}
   \frac{\partial^2 Z[\Phi_S]}{\partial\Phi_S(x)\partial\Phi_S(x')}
\!\!=\!\!\frac{1}{(2\pi)^4}\!\!\!\int_{-\infty}^{+\infty}\!\!\!\!\!d^4p\,
  \,e^{ip(x'-x)}\!\!\sum_{n=0}^{\infty}\!\!\! a^3(z_{\text{min}})
  \frac{\partial_z\h{\Phi}^{\star}_n(z_{\text{min}})\h{\Phi}_n(z_{\text{min}})}
  {p^2+m^2_n-i\epsilon}.\! \!\!\!\!\!
  \eq
From a minimal extension of the action (\ref{free_action}) though
a dynamical $U(1)$ bulk field coupled to $\Phi$
\beq
S_{\text{{AdS}}}[\Phi,\Phi^\star ,A^m]=\int d^4xdz\ \sqrt{-g}
\left(-\frac{1}{4}F^{mn}F_{mn}
  + e^{-\chi}\left((D^m\Phi)^\star D_m\Phi+m_S^2 \, \Phi^\star \Phi\right)\right)\nonumber\,,
\eq
with $D_n \Phi= \partial_n \Phi -i e A_n \Phi$, the  $n-$point correlation functions are obtained by 
the correspondence
\begin{eqnarray}
\label{Z_cor}
{Z_{\text{{QCD}}}({\stmath c,\bar{c},n+\bar{n}})} &=& { \left\langle\exp\left(\int d^4x\ 
({\stmath n_\mu+\bar{n}_\mu})J^\mu+{ \bar{c}}\,O+{\stmath c}\,O^\dagger\right)\right\rangle_{\text{{QCD}}}} \\
&=& \exp\left(-S^{cl}_{\text{{AdS}}}
[\Phi({\stmath c}),\Phi^\star({\stmath\bar{c}}),A^m({\stmath n_\mu+\bar{n}_\mu})]
\right) 
\end{eqnarray}
where
${\stmath c}, {\stmath\bar{c}}, {\stmath n+\bar{n}}$ are
 4d sources for QCD which appear as boundary
conditions for the 5d 
bulk fields.
Correlation functions of QCD operators are obtained by an expansion
 to linear-order with respect to the sources.
QCD operators are coupled to asymptotic states which are boundary conditions
of {\it \structure{normalizable}} bulk fields $\Phi^{(0)},$ $\Phi^{*(0)}$ (scalar probe).
The electromagnetic current is dual to a massless {\it non-normalizable} 5d vector field $A_m(x,z)$ which satisfies Maxwell equation 
  $ \partial^m(z^{-1}F_{mn}) = 0 \,,\forall n\,.$ Plane-wave solutions read (with the condition $\lim\limits_{z\to 0} \, A(z)=1$)
\beq
   A_{\mu}(x,z) = {\stmath n_{\mu}} \, e^{iq\cdot x}A(z)\,, \quad
    A_z(x,z) = e^{iq\cdot x}A_0(z)\,,\quad n^2=1 
\eq
in the Lorentz-like gauge, with
$A_0(z) = -i(q\cdot n)/q^2\partial_zA(z) $ and $A(z)=Q \, z \, K_1(Qz)\,\,(Q^2=q\cdot q)\,.$
%
%
%
%

Evaluating the $ \gamma^{(*)} A \to \gamma^{(*)} A'$ {\aut Compton} scattering amplitude
on a spinless, or spin-averaged target
 requires to study a 4-point function, the first non-trivial correlator involving the propagation inside the bulk\footnote{Deep Inelastic Scattering only involves 3-point functions when evaluating matrix elements of the $U(1)$ current on a given set of intermediate states in AdS.}.
Based on (\ref{Z_cor}), 
one should  extract the coefficient of $\bar{c} \, n_\mu \, \bar{n}_\nu \, c\,.$
This is done by solving iteratively the coupled classical equations in terms of the free 
bulk fields $A^{(0)},$  $\Phi^{*(0)},$ $\Phi^{(0)}\,.$
One then deduces the classical action, up to $e^2.$
The boundary condition at $z=0$ for $A^{(0)},$  $\Phi^{*(0)},$ $\Phi^{(0)}$ are $n_{\mu}+\bar{n}_{\mu},$ $c,$ and $\bar{c}\,,$ 
which enter 
 {\it in a linear
way} in these free fields: the
$\bar{c}n_\mu\bar{n}_\nu c$ and 
$A^{(0)}\,A^{(0)} \,\Phi^{*(0)} \, \Phi^{(0)}$ coefficients are equal. 
Contracting  the result on the physical plane \nolinebreak wave  boundary conditions 
for $A^{(0)}$ (non-normalizable),~$\Phi^{*(0)},$ $\Phi^{(0)}$~(normalizable)~gives~the~contributions:\linebreak 
\beq
\label{diagrams}
\begin{array}{ccc}
\psfrag{q1}{}
\psfrag{q2}{}
\psfrag{p1}{}
\psfrag{p2}{}
\psfrag{s}{}
\vspace{-.1cm}
\scalebox{1}{\raisebox{-0.46 \totalheight}{\epsfig{file=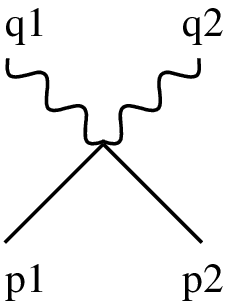,width=1cm}}}
&
\hspace{2cm}
\psfrag{q1}{}
\psfrag{q2}{}
\psfrag{p1}{}
\psfrag{p2}{}
\psfrag{s}{}
\scalebox{1}{\raisebox{-0.46 \totalheight}{\epsfig{file=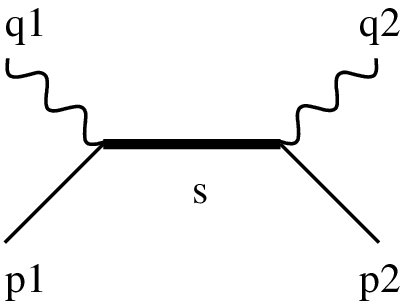,width=1.5cm}}}
&
\hspace{2cm}
 \psfrag{q1}{}
\psfrag{q2}{}
\psfrag{p1}{}
\psfrag{p2}{}
\psfrag{s}{}
\scalebox{1}{\raisebox{-0.46 \totalheight}{\epsfig{file=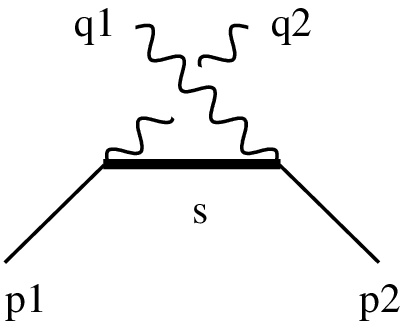,width=1.5cm}}} \\
\text{contact term}
&
\hspace{2cm}\text{$s$-channel diagram}
&
\hspace{2cm}\text{$u$-channel diagram}
\end{array}
\eq

\section{Compton scattering off an unpolarized target $A$
$    \gamma^{\star}(q_1) + P(p_1) \rightarrow \gamma^{\star}(q_2) + P(p_2)$}

Choosing $q_1$, $q_2$ and $p=p_1+p_2$ as the three independent momenta, the 
10 independent parity-conserving tensors of rank 2 reduce to 5 after implementing the 
electromagnetic gauge invariance:
\begin{eqnarray}
\label{amplitude_CS}
 \hspace{-.3cm} T^{\mu\nu} &=& \alert{ V_1} {\violet\left(g^{\mu\nu} - \frac{q_1^{\mu}q_1^{\nu}}{q_1^2} 
    - \frac{q_2^{\mu}q_2^{\nu}}{q_2^2}
    + q_1^{\mu}q_2^{\nu}\frac{(q_1.q_2)}{q_1^2q_2^2} \right)} + \alert{ V_2} {\violet\left( p^{\mu} - q_1^{\mu}\frac{(p.q_1)}{q_1^2}\right)
   \left(p^{\nu} - q_2^{\nu}\frac{(p.q_2)}{q_2^2}\right)} \nonumber \\
   &+& \alert{ V_3} {\violet\left(q_2^{\mu} - q_1^{\mu}\frac{(q_1.q_2)}{q_1^2}\right)
   \left(q_1^{\nu} - q_2^{\nu}\frac{(q_1.q_2)}{q_2^2}\right)} + \alert{ V_4} {\violet\left( p^{\mu} - q_1^{\mu}\frac{(p.q_1)}{q_1^2}\right)
   \left(q_1^{\nu} - q_2^{\nu}\frac{(q_1.q_2)}{q_2^2}\right)} \nonumber\\
   &+& \alert{ V_5} {\violet\left(q_2^{\mu} - q_1^{\mu}\frac{(q_1.q_2)}{q_1^2}\right)
   \left(p^{\nu} - q_2^{\nu}\frac{(p.q_2)}{q_2^2}\right)} 
   = \sum_{i=1}^5 \alert{V_i}({\stmath p_1^2,p_2^2,q_1^2,q_2^2,s,u})\,{\violet \V_i^{\mu\nu}(p,q_1,q_2)}
   \,. \,\,
\end{eqnarray}
The DVCS amplitude reduces to the Virtual Compton Scattering amplitude (VCS) when the produced photon is real, and is expressed in terms of 3 independent form factors. When additionally the incoming photon is real, the  Real Compton Scattering amplitude depends only of 2 form factors.

One can now rely on the AdS/QCD correspondence to evaluate these amplitudes.
In the on-shell limit ($p_1^2=p_2^2=-m^2$), the gauge-invariant amplitude (\ref{amplitude_CS}) reduces to 
\begin{eqnarray}
\label{Tmunu}
        T^{\mu\nu} &=&  e^2\left(2\, \alert{\C_1}\,\V_1^{\mu\nu} 
          +  \alert{\C_+} \left(\V_2^{\mu\nu} + \V_3^{\mu\nu}\right)
          +  \alert{\C_-} \left(\V_4^{\mu\nu} + \V_5^{\mu\nu}\right)\right) \nonumber \\
\C_{\pm}(m^2,q_1^2,q_2^2,s,u) &=& \F_1(m^2,q_1^2,q_2^2,s) \pm 
      \F_1(m^2,q_2^2,q_1^2,u) \,,
\end{eqnarray}
and is thus expressed in terms of
only 3 Compton form factor out of a possible 5.
The form factors
$\F_1(m^2,q_1^2,q_2^2,k^2)$ and
    $\C_1(m^2,q_1^2,q_2^2)$ obtained from the AdS/QCD correspondence read (similar results were obtained within the hard wall model in Ref.~\cite{GX09})
\begin{eqnarray}
\label{F1}
    \hspace{-1cm}    \F_1\left(m^2,q^2_1,q_2^2,k^2\right)\!\! &\!\!=\!\! & \!
      \! \!\!\! \iint \! dz_1\,dz_2\,\,z_1^{-3}e^{-\chi(z_1)}\,A_1(z_1)\,\h{\Phi}_i(z_1)\,
        \h{G}_0(z_1,z_2,k^2)
        \,\h{\Phi}_f^{\star}(z_2)\,A_2^{\star}(z_2)\,z_2^{-3}e^{-\chi(z_2)},\,\,\,\\
    \label{C1}
     \C_1(m^2,q_1^2,q_2^2) &=& \int dz\,z^{-3}e^{-\chi(z)}\,A_1(z)\,A_2^{\star}(z)\,\h{\Phi}_i(z)\,\h{\Phi}_f^{\star}(z) \,.
\end{eqnarray}
The $\F_1$ form factor has a natural {\aut K\"allen-Lehman} spectral representation 
\begin{eqnarray}
\label{F1KL}
     \F_1\left(m^2,q^2_1,q_2^2,k^2\right) &=&  \sum_{n=0}^{\infty}
    \frac{{\stmath \Gamma(m^2,m_n^2,q_1^2)}\,{\stmath\Gamma^\star(m^2,m_n^2,q_2^2)}}
    {k^2+m^2_n-i\epsilon} \,,
\\
\label{Gamma}
    {\stmath \Gamma}(m^2,m_n^2,Q^2) &=& Q\int dz\,z^{-2}e^{-\chi(z)}
    K_1(Qz)\,\h{\Phi}_m(z)\,\h{\Phi}^*_n(z) \,.
\end{eqnarray}
The vertex function $\Gamma$
reduces to the electromagnetic form factor when $m^2=m_n^2\,.$ Starting from the on-shell Compton amplitude (\ref{Tmunu}),
its absorptive part in the forward limit
\beq
\text{Im}\,T^{\mu\nu}(q^2,s) \approx 
  e^2\left.\left(\px{m_n^2}{n}\right)^{-1}\right|_{m_n^2=-s}
  \left|{\stmath\Gamma(m^2,-s,q^2)}\right|^2
 \alert{ \left(p^{\mu} + \frac{1}{x}q^{\mu}\right)
  \left(p^{\nu} + \frac{1}{x}q^{\nu}\right)} 
\eq
gives the hadronic tensor of Deep Inelastic Scattering (DIS),  showing that only \alert{$F_L$} survives, in contradiction with Callan-Gross relation.
The scaling properties 
are governed by $\Gamma(m^2,m_n^2,Q^2)$:
\beq
  \Gamma(m^2,m_n^2,Q^2) \approx \int_0^{1/Q}\frac{dz}{z^3}\,e^{-\chi(z)}
    \,\h{\Phi}(z)\,\h{\Phi}^*_n(z) \,.
\eq
For $Q^2\gg m^2,\ Q^2\gtrsim m_n^2$, the scaling of the
hadron state $\phi(z)\underset{z\rightarrow 0}{\sim} z^{\Delta}$ ($\Delta$ is the scaling dimension of the operator creating the spinless hadron) leads to
%
$
F_2 (Q^2,x) \propto (1/Q^2)^{\Delta-1}
$
calling for \  $\Delta=1\,,$ while for
$Q^2\gg m^2,\, m_n^2$, the scaling properties
$\phi(z)\underset{z\rightarrow 0}{\sim} z^{\Delta}$ and
    $\phi_n(z)\underset{z\rightarrow 0}{\sim} z^{\Delta}$ lead to
$
F_{\gamma}(Q^2) \propto (1/Q^2)^{\Delta-1}
$\nolinebreak
calling  for $\Delta=2$ \cite{BRO07}.
This generic AdS/QCD tension due to the same scaling obtained for structure function and form factors is
 impossible to reconcile with a partonic picture in a simple  way \cite{PRSW}.

The VCS amplitude for $\gamma^* A \to \gamma A'$ has the same tensorial structure as point-like scalar electrodynamic and depends on a single electromagnetic form factor $\C_1(m^2,q_1^2,0)$ (\ref{C1}).
In perturbative QCD, these form factors can in principle be related,
through factorization, to \structure{generalized parton distributions (GPDs)}.
The partonic interpretation is based on the convolution of \structure{real GPDs} with
coefficient functions which contain both a real and an imaginary part. In contrast,
the above holographic DVCS amplitude  has no absorptive part, while it is seen  experimentally (through interference with the Bethe-Heitler amplitude). This may call 
for stringy corrections.
Besides, the asymptotic behavior in $Q^2$ of the holographic DVCS cross-section is governed by the power-law behavior of the electromagnetic form factor, while it is expected to scale like DIS amplitude. 

The Real Compton scattering ($q_1^2= q_2^2= 0$) amplitude provides informations on electric and magnetic static polarizabilities $\alpha_{E}$ and $\alpha_{M}$,  defined as quadratic coefficients (in the photons energies) in
corrections to {\aut Thomson}
scattering,
defined in lab-frame $\overrightarrow{p_1}=\vec{0}$ ($q_i=\omega_i(1,\h{q}_i), \, \omega_i^2\ll m_{\pi}^2$)
\beq
A(\gamma\pi\rightarrow\gamma\pi) = 2\, e^2 \overrightarrow{\epsilon_1}
\cdot\overrightarrow{\epsilon_2}
    + 8\pi m_{\pi}\,\omega_1\omega_2
    \bigl({\stmath \alpha_{E}}\overrightarrow{\epsilon_1}\cdot\overrightarrow{\epsilon_2} + {\stmath\beta_{M}}\,
  (\overrightarrow{\epsilon_1}\times \h{q}_1)\cdot(\overrightarrow{\epsilon_2}\times \h{q}_2)\bigr)\,.
\eq
%
 The holographic Compton amplitude leads to vanishing static polarizabilities $\alpha_{E}$ and $\alpha_{M}$, 
in contradiction with the most recent experimental values for the pion
\cite{EXP}.
%
One way to circumvent this problem would be to consider a non-minimal coupling between the bulk field $\Phi$ and the $U(1)$ field.

To conclude, AdS/QCD Compton amplitude is trivial in the low-energy limit. It has
 \alert{no partonic interpretation} in the high-energy limit.
Most popular AdS/QCD models incorporating flavor symmetry do not cure all the problems.
\vspace{.15cm}

Work supported in part by  the grant ANR-06-JCJC-0084.


\end{document}